\documentclass[cmt]{svjour}
\usepackage{graphics}
\begin{document}
\title{Thermodynamic stability conditions for nonadditive 
composable entropies}
\author{Wada Tatsuaki\inst{1} 
}                     
\institute{Department of Electrical and Electronic Engineering, 
Ibaraki University, Hitachi,~Ibaraki, 316-8511, Japan}
\date{Received: date / Accepted: date}
%
\maketitle
\begin{abstract}
The thermodynamic stability conditions (TSC) of nonadditive and 
composable entropies are discussed. Generally the concavity 
of a nonadditive entropy with respect to internal energy is 
not necessarily equivalent to the corresponding TSC. 
It is shown that both the TSC of 
Tsallis' entropy and that of the $\kappa$-generalized Boltzmann entropy
are equivalent to the positivity of the standard heat capacity.
\end{abstract}
\section{Introduction}
\label{intro}
It is well known that the conventional thermodynamic stability
condition (TSC) \cite{Callen} is the concavity of the Boltzmann-Gibbs (BG) 
entropy $S^{\rm BG}$ with respect to internal energy $U$, i.e.,
\begin{equation}
    \frac{\partial^2 S^{\rm BG}}{\partial U^2} < 0,
    \label{BG-concavity}
\end{equation}
which is also equivalent to the positivity 
of heat capacity $C \equiv \partial U / \partial T$,
since the temperature $T$ is defined by
\begin{equation}
    \frac{1}{T} \equiv \frac{\partial S^{\rm BG}}{\partial U},
\end{equation}
and
\begin{equation}
    \frac{1}{C} = \left\{ \frac{\partial T}{\partial (1/T)} \right\} 
            \left\{ \frac{\partial (1/T)}{\partial U} \right\} =
        -T^2 \frac{\partial^2 S^{\rm BG}}{\partial U^2}.
\end{equation}

The conventional thermodynamic stability arguments are
based on both the extremum principle and additivity of
the BG entropy.
Nowadays nonadditive entropies have been extensively considered
in the literature since the generalized formalism
based on Tsallis' nonadditive entropy \cite{Tsallis88,NEXT2001}
has been successfully
applied to a variety of complex systems. 
The Tsallis entropy is a generalization of the BG entropy by
one real parameter of $q$,
\begin{equation}
  S_q \equiv \frac{\sum_i p_i^q - 1}{1-q},
\end{equation}
where $p_i$ stands for a probability of $i$-th state.
For the sake of simplicity, we set the Boltzmann constant to unity
throughout this paper. $S_q$ reduces to the BG entropy in the limit 
of $q \to 1$.

During the early stages of the development of Tsallis' thermostatistics,
Ramshaw \cite{Ramshaw95} pointed out that 
the concavity of Tsallis' entropy with respect to the
standard expectation energy $U_1 \equiv \sum_i p_i E_i$ 
is not sufficient to guarantee the thermodynamic stability
for all values of $q$ because $S_q$ is nonadditive.
He derived the TSC of the nonadditive $S_q$
from the TSC of the additive R\'enyi entropy \cite{Renyi}
\begin{equation}
  S_q^{\rm R} \equiv \frac{\ln \sum_i p_i^q}{1-q},
\end{equation}
by utilizing the nonlinear functional relation between both entropies
\begin{equation}
   S_q^{\rm R} = \frac{ \ln [1+(1-q)S_q]}{1-q}.
   \label{T-transform}
\end{equation}
The present author \cite{Wada02} further studied the TSC of $S_q$ 
by directly utilizing the nonadditivity Eq. (\ref{pseudo-additivity}) 
of $S_q$, and shown that the resultant
TSC is equivalent to the positivity of the standard heat capacity $C$.

In this paper we discuss the TSCs of nonadditive composable entropies,
and show that the TSC of such a nonadditive entropy is not equivalent 
to the concavity of the entropy with respect to internal energy. 
In the next section, after the explanation of the concept
of {\it composability}, we briefly review the standard TSC of
an additive entropy. We then discuss the TSCs of two different kinds
of nonadditive composable entropies. The subsection 2.1 deals with 
the TSC of the Tsallis $q$-entropy, and in the subsection 2.2 we consider
the TSC of the $\kappa$-generalized Boltzmann entropy. It is shown that
the TSCs of the both nonadditive entropies are equivalent to the positivity
of the standard heat capacity.
The final section is devoted to the conclusions.

\section{Thermodynamic stability conditions of composable entropies}
\label{sec:1}
The concept of {\it composability} \cite{Hotta99} is important and
quite useful when we consider a thermodynamic system which
consists of independent subsystems. Suppose a total system
consists of two independent subsystems $A$ and $B$.
If the total entropy of any sort ${\mathcal S}(A, B)$ is 
a bivariate and symmetric function $f$ of the subsystem entropies 
${\mathcal S}(A)$ and ${\mathcal S}(B)$, i.e.,
\begin{equation}
   {\mathcal S}(A, B) = f({\mathcal S}(A), {\mathcal S}(B)) 
                      = f({\mathcal S}(B), {\mathcal S}(A)),
  \label{composability}
\end{equation}
we say ${\mathcal S}$ is {\it composable}.
This means that the total entropy ${\mathcal S}(A, B)$ can be 
built of just the two macroscopic quantities, ${\mathcal S}(A)$ and 
${\mathcal S}(B)$.
Hence we can further discuss the thermodynamic properties of any composable
system without knowing its underlying microscopic dynamics.
For example, the zeroth law of thermodynamics 
\cite{Abe01,Martinez01,Abe01-Heat,Wang03} 
for Tsallis' entropy has been discussed by assuming the composability.
It is well known that the BG entropy is composable and additive,
\begin{equation}
   S^{\rm BG}(A, B) = S^{\rm BG}(A) + S^{\rm BG}(B).
\end{equation}
Tsallis' entropy is also composable but it's nonadditive, i.e., 
so-called pseudo-additive \cite{Tsallis88,NEXT2001}
\begin{equation}
   S_q(A, B) = S_q(A) + S_q(B) + (1-q) S_q(A) S_q(B).
   \label{pseudo-additivity}
\end{equation}

Let us now review the TSC of any additive entropy $S(U)$.
We denote the maximum entropy of a thermodynamic system with 
an internal energy $U$ by $S(U)$. For the sake of simplicity,
the internal energy is assumed to be additive, i.e., $U(A, B) = U(A)+U(B)$,
in this paper.
It is however worth noting that
Wang \cite{Wang03} has discussed the zeroth and first laws of thermodynamics 
by utilizing nonadditive energy expectation value within the framework of 
nonextensive statistical mechanics.
 
The essence of conventional thermodynamic stability \cite{Callen} lies in
the entropy maximum principle and additivity of $S(U)$. 
It is known that the relation between the concavity of $S(U)$ and
the standard TSC is straightforward as follows. If one
transfer an amount of energy $\Delta U$ from one of two identical
subsystems to the other subsystem, the total entropy
changes from its initial value of $S(U,U)=2 S(U)$ to 
$S(U+\Delta U, U-\Delta U) = S(U+\Delta U) + S(U-\Delta U)$. 
The entropy maximum principle demands that the
resultant value of the entropy is not larger than the initial one, i.e.,
\begin{equation}
   2 S(U) \ge S(U+\Delta U) + S(U-\Delta U),
  \label{TSC}
\end{equation}
which is the TSC of the additive entropy $S(U)$.  In the limit of
$\Delta U \to 0$, Eq. (\ref{TSC}) reduces to the concavity of $S(U)$
with respect to $U$,
\begin{equation}
   \frac{\partial^2 S(U)}{\partial U^2} \le 0.
  \label{concavity}
\end{equation}
Thus the TSC and the concavity of $S(U)$ with respect to $U$ are equivalent 
each other when $S(U)$ is additive. 
However this is not the case for a nonadditive entropy 
as we will see in the subsequent subsections.

\subsection{Tsallis' entropy}
Let us now turn focus on the TSC of Tsallis' entropy \cite{Wada02}.
Taking account of the pseudo-additivity
Eq. (\ref{pseudo-additivity}), the TSC of $S_q$ can be written by
\begin{eqnarray}
   2 S_q(U) &+& (1-q) \left[ S_q(U)\right ]^2 \nonumber \\
   & \ge &  S_q(U + \Delta U) + S_q(U - \Delta U)
        + (1-q) S_q(U + \Delta U) S_q(U - \Delta U).
  \label{TSC-Sq}
\end{eqnarray}
The physical meaning of this inequality is same as that of the
conventional TSC Eq. (\ref{TSC}). If we transfer a small amount of
energy $\Delta U$ from one of two identical subsystems
to the other subsystem, then the total entropy changes from its
initial value $S_q(U, U)$ of the left-hand-side to 
the resultant value $S_q(U + \Delta U, U - \Delta U)$ of
the right-hand-side in Eq. (\ref{TSC-Sq}).  
The principle of the maximum Tsallis entropy 
demands that the resultant entropy should not be larger than the initial one.
The remarkable difference between
Eq. (\ref{TSC-Sq}) and Eq. (\ref{TSC}) is the presence of the nonlinear
term proportional to $1-q$, which originally arises from the
pseudo-additivity of Eq. (\ref{pseudo-additivity}).

In the limit of $\Delta U \to 0$, Eq. (\ref{TSC-Sq}) reduces to 
the differential form
\begin{equation}
   \frac{\partial^2 S_q(U)}{\partial U^2} + (1-q) \left\{
     S_q(U) \frac{\partial^2 S_q(U)} {\partial U^2} -
     \left( \frac{\partial S_q(U)}{\partial U} \right)^2
     \right\} \le 0.
  \label{diff-TSC-Sq}
\end{equation}
Now we introduce the $q$-generalized temperature $T_q$ defined by
\begin{equation}
  \frac{1}{T_q} \equiv \frac{\partial S_q}{\partial U},
\end{equation}
and the $q$-generalized heat capacity $C_q$ defined by
\begin{equation}
  \frac{1}{C_q} \equiv \frac{\partial T_q}{\partial U} 
      = -T_q^2 \frac{\partial^2 S_q}{\partial U^2}.
\end{equation}
Then Eq. (\ref{diff-TSC-Sq}) becomes
\begin{equation}
   \frac{1 + (1-q)S_q}{C_q} + (1-q) \ge 0.
   \label{TSC-Cq-Sq}
\end{equation}
We thus see that only the positivity of $C_q$, or equivalently
the concavity of $S_q(U)$, is not
sufficient to satisfy the TSC of the nonadditive Tsallis entropy. 
When $q \le 1$, however, the positivity of $C_q$ guarantees 
to satisfy Eq. (\ref{TSC-Cq-Sq}) since $1+(1-q)S_q =
\sum_i p_i^q$ is always positive.

Next we shall consider the relation between the TSC and
the positivity of the standard heat capacity.
By maximizing the total Tsallis entropy $\delta S_q(A, B)=0$
under the constraint of the total energy conservation $\delta U(A, B) = 0$,
we obtain the equilibrium condition for a composite system
described by Tsallis' entropy,
\begin{equation}
   T_q(A) \{ 1+(1-q)S_q(A) \}
    =  T_q(B) \{ 1+(1-q)S_q(B) \},
\end{equation}
which should be equal to the intensive temperature $T$
of the total system \cite{Abe01}, i.e.,
\begin{equation}
   T = \{1+(1-q)S_q \} \cdot T_q.
   \label{T-Tq}
\end{equation}
By differentiating the both sides of Eq.(\ref{T-Tq}) with respect to $U$, 
we obtain
the relation between the standard heat capacity 
$C \equiv \partial U/\partial T$ and 
the $q$-generalized heat capacity $C_q$ as
\begin{equation}
   \frac{1}{C} = \frac{1+(1-q)S_q}{C_q} + 1-q.
\end{equation}
The TSC condition of Eq. (\ref{TSC-Cq-Sq}) is thus 
equivalent to the positivity of the conventional heat capacity,
\begin{equation}
  C \ge 0.
  \label{TSC-C}
\end{equation}

\subsection{The $\kappa$-generalized Boltzmann entropy}

Next we turn focus on another nonadditive composable entropy.
Kaniadakis \cite{Giorgio01,Giorgio-Antonio02} has introduced another type 
of one parameter deformations of the exponential and logarithmic functions
\begin{eqnarray}
   \exp_{\{ \kappa \}} (x) \equiv \left[ \sqrt{1+\kappa^2 x^2} + \kappa x 
             \right]^{1/\kappa}, \\
   \ln_{\{ \kappa \}}(x) \equiv \frac{x^{\kappa}-x^{-\kappa}}{2 \kappa},
   \label{kappa-log}
\end{eqnarray}
where $\kappa$ is a real parameter with $-1 < \kappa < 1$. 
In the limit of $\kappa \to 0$, the $\exp_{\{ \kappa \}}(x)$ and 
$\ln_{\{ \kappa \}}(x)$
reduce to the standard exponential and logarithmic functions, respectively. 
The $\kappa$-deformed exponential and logarithmic functions
have the following properties
\begin{eqnarray}
   \exp_{\{ \kappa \}}(x) \; \exp_{\{ \kappa \}}(y) = 
   \exp_{\{ \kappa \}}( x \stackrel{\kappa}{\oplus} y),  \\
   \ln_{\{ \kappa \}}(x y) = \ln_{\{ \kappa \}}(x) \stackrel{\kappa}{\oplus}
                  \ln_{\{ \kappa \}}(y),
\end{eqnarray}
where $\kappa$-sum is defined by
\begin{equation}
   x \stackrel{\kappa}{\oplus} y
   \equiv x \sqrt{1+\kappa^2 y^2} + y \sqrt{1+\kappa^2 x^2}.
   \label{kappa-sum}
\end{equation}
Kaniadakis has defined the $\kappa$-deformed entropy as
\begin{equation}
   S_{\kappa}^{\rm K} \equiv \sum_i p_i \ln_{\{ \kappa \}} p_i,
\end{equation}
to construct a generalized statistical mechanics 
in the context of special relativity \cite{Giorgio02}.
$S_{\kappa}^{\rm K}$ reduces to the Shannon-BG entropy in the limit of 
$\kappa \to 0$.
The $\kappa$-additivity defined by Eq. (\ref{kappa-sum}) is the additivity
of relativistic momenta as shown in Ref. \cite{Giorgio02}.

We shall here consider the TSC of the $\kappa$-generalized Boltzmann entropy
defined by
\begin{equation}
   S_{\kappa}^{\rm B}(U) \equiv \ln_{\{ \kappa \}} W(U),
   \label{kappa-entropy}
\end{equation}
where $W$ is the number of microcanonical configurations of
a thermodynamic system. $S_{0}^{\rm B}(U)=\ln W(U)$ is
the standard Boltzmann entropy.
Note that this $S_{\kappa}^{\rm B}$ is different from 
the Kaniadakis $\kappa$-entropy $S_{\kappa}^{\rm K}$, which
is an entropy {\it \`a la} Gibbs,
whereas the $S_{\kappa}^{\rm B}$ is one {\it \`a la} Boltzmann.
The reason of using $S_{\kappa}^{\rm B}$, instead of $S_{\kappa}^{\rm K}$,
is that it has the composability (the $\kappa$-additivity),
\begin{equation}
  S_{\kappa}^{\rm B}(A, B) = S_{\kappa}^{\rm B}(A) \stackrel{\kappa}{\oplus}
                      S_{\kappa}^{\rm B}(B). 
\end{equation}
By utilizing this composition rule we readily write down
the expression of the TSC of $S_{\kappa}^{\rm B}$, 
\begin{eqnarray}
  S_{\kappa}^{\rm B}(U, U) &\ge& 
 S_{\kappa}^{\rm B}(U+\Delta U, U-\Delta U), \nonumber \\
       2 S_{\kappa}^{\rm B}(U) \sqrt{1+\kappa^2 [S_{\kappa}^{\rm B}(U)]^2} 
    &\ge& S_{\kappa}^{\rm B}(U+\Delta U) 
                 \sqrt{1+\kappa^2 [S_{\kappa}^{\rm B}(U-\Delta U) ]^2}
    \nonumber \\
     &&+ S_{\kappa}^{\rm B}(U-\Delta U) 
                 \sqrt{1+\kappa^2 [S_{\kappa}^{\rm B}(U+\Delta U) ]^2}.
    \label{TSC-S_kappa}
\end{eqnarray}
After some algebra, in the limit of $\Delta U \to 0$,  Eq. (\ref{TSC-S_kappa})
becomes 
\begin{equation}
  \frac{ 1+2 \kappa^2 [ S_{\kappa}^{\rm B}(U) ]^2 }
       { 1+\kappa^2 [ S_{\kappa}^{\rm B}(U) ]^2 }
\sqrt{ 1+\kappa^2 [ S_{\kappa}^{\rm B}(U) ]^2 }
  \left\{
      \frac{ \partial^2 S_{\kappa}^{\rm B}(U) }{ \partial U^2 }
         - \left( \frac{ \partial S_{\kappa}^{\rm B}(U) }{\partial U} 
           \right)^2
           \frac{ \kappa^2 S_{\kappa}^{\rm B}(U) }
                { 1+\kappa^2 [S_{\kappa}^{\rm B}(U) ]^2 } 
  \right\}
   \le 0.
\end{equation}
Since the terms in front of the brace is always positive, the differential
form of the TSC of $S_{\kappa}^{\rm B}$ becomes
\begin{equation}
      \frac{ \partial^2 S_{\kappa}^{\rm B} }{ \partial U^2 }
         - \left( \frac{ \partial S_{\kappa}^{\rm B} }{\partial U} 
           \right)^2
           \frac{ \kappa^2 S_{\kappa}^{\rm B} }
                { 1+\kappa^2 [S_{\kappa}^{\rm B}]^2 } 
   \le 0.
   \label{diff-TSC-Sk}
\end{equation}
Furthermore since $W(U)$ is much larger than unity for almost every energy $U$
in conventional thermodynamics situations, 
$S_{\kappa}^{\rm B}(U)=\ln_{\{ \kappa \}} W(U)$ 
is positive for such an energy.
Thus the concavity of $S_{\kappa}^{\rm B}(U)$ guarantees to satisfy the TSC 
of Eq. (\ref{diff-TSC-Sk}) unlike the case of the TSC of Tsallis' entropy.
Interestingly the concavity 
$\partial^2 S_{\kappa}^{\rm B}(U) / \partial^2 U < 0$ is the sufficient, but
not necessary, condition in order to satisfy the TSC of $S_{\kappa}^{\rm B}$.
For example, it is possible to satisfy Eq. (\ref{diff-TSC-Sk}) 
even if $S_{\kappa}^{\rm B}$ is convex (or $C_{\{ \kappa \}}$ is negative).

Next we turn to the relation between the TSC and the standard heat capacity.
We introduce the $\kappa$-generalized temperature $T_{\{ \kappa \}}$ defined by
\begin{equation}
   \frac{1}{T_{\{ \kappa \}}} \equiv 
       \frac{\partial S_{\kappa}^{\rm B}(U)}{\partial U}.
   \label{T_kappa}
\end{equation}
Following the similar arguments of the equilibrium condition
(or thermodynamic zeroth law) for Tsallis' entropy in the previous subsection,
we obtain the relation between $T_{\{ \kappa \}}$
and the corresponding intensive temperature $T=T_{\{0\}}$ as
\begin{equation}
  T = T_{\{ \kappa \}} \sqrt{ 1+\kappa^2 [ S_{\kappa}^{\rm B}(U) ]^2 }
  \label{T-T_kappa}
\end{equation}
Alternatively we can obtain this relation from the definition 
Eq. (\ref{T_kappa}) of $T_{\{ \kappa \}}$,
\begin{equation}
 \frac{1}{ T_{\{ \kappa \} }} 
   = \frac{\partial}{\partial U} \left( 
           \frac{ W^{\kappa}-W^{-\kappa} }{2 \kappa}
      \right) =
\left( \frac{\partial \ln W}{\partial U} \right)
\left( \frac{W^{\kappa}+W^{-\kappa}}{2 \kappa} \right)= 
 \frac{1}{ T_{\{ 0 \} }} \sqrt{ 1+\kappa^2  [S_{\kappa}^{\rm B}]^2 }.
\end{equation}
By differentiating the both sides of Eq.(\ref{T-T_kappa}) with respect to $U$, 
we obtain the relation between the standard heat capacity $C$ and  
the $\kappa$-generalized heat capacity 
$C_{\{ \kappa \}} \equiv \partial U/\partial T_{\{ \kappa \}}$ as follows.
\begin{eqnarray}
   \frac{1}{C} &=& \sqrt{1+\kappa^2 [S_{\kappa}^{\rm B}]^2} 
     \left\{ \frac{1}{C_{\{ \kappa \}}} + \frac{\kappa^2 S_{\kappa}^{\rm B}}
               {1+\kappa^2 [S_{\kappa}^{\rm B}]^2}
         \right\} \nonumber \\
   &=& -T_{\{\kappa \}}^2 \sqrt{1+\kappa^2 [S_{\kappa}^{\rm B}]^2} 
    \left\{
      \frac{ \partial^2 S_{\kappa}^{\rm B} }{ \partial U^2 }
         - \left( \frac{ \partial S_{\kappa}^{\rm B} }{\partial U} 
           \right)^2
           \frac{ \kappa^2 S_{\kappa}^{\rm B} }
                { 1+\kappa^2 [S_{\kappa}^{\rm B}]^2 }
    \right\},
  \label{C}
\end{eqnarray}
where we used Eq. (\ref{T_kappa}) and the relation 
$1/C_{\{ \kappa \}}= -T_{\{\kappa \}}^2 \cdot
\partial^2 S_{\kappa}^{\rm B} / \partial U^2 $. 
By comparing Eq (\ref{C}) with Eq. (\ref{diff-TSC-Sk}), we finally find that 
the TSC of $S_{\kappa}^{\rm B}$ is also equivalent to the positivity of 
standard heat capacity $C \ge 0$.

\section{Conclusions}
We have discussed the thermodynamic stability conditions (TSC) 
of the two nonadditive composable entropies, which are the Tsallis 
and $\kappa$-generalized Boltzmann entropies.
Unlike the TSC of an additive entropy, the concavity 
of a nonadditive composable entropy is not necessarily equivalent to the 
corresponding TSC in general.
It is shown that in addition to the TSC of 
Tsallis' entropy, the TSC of the $\kappa$-generalized Boltzmann entropy
is also equivalent to the positivity of the standard heat capacity.
It is interesting to further study whether it is common feature
to composable entropy. 

Nonadditive entropy can provide a basic 
ingredient in order to describe non-separable or correlated systems whose
thermodynamic behaviors are complex and anomalous.
It is worth noting that the we have derived the TSCs with
resort to the composability of the nonadditive entropies. 
However composability of systems restricts our consideration
to separable systems, which are divisible into independent subsystems.
For non-composable entropies we have not yet known the expression
of the corresponding TSCs.
 
%

\section*{Acknowledgments}
The author greatly thanks Prof. M. Sugiyama for giving 
the opportunity to reconsider the thermodynamic stability
of the nonextensive entropies. 
He also thanks G. Kaniadakis and S. Abe for valuable discussions.



\begin{thebibliography}{}
%
%
\bibitem{Callen}
Callen HB \textit{Thermodynamics and an Introduction to Thermostatics}
2nd ed. (Wiley New York 1985) Chap. 8

\bibitem{Tsallis88}
Tsallis C (1988) Possible generalization of Boltzmann-Gibbs statistics.
J. Stat. Phys. \textbf{52} 479

\bibitem{NEXT2001}
Kaniadakis G, Lissia M, Rapisarda A (editors) (2001)
\textit{Proc. of the international school and workshop on non extensive
thermodynamics and physical applications (NEXT2001)}
Physica A \textbf{305} Nos. 1-2

\bibitem{Ramshaw95}
Ramshaw JD (1995) Thermodynamic stability conditions for the Tsallis 
and R\'enyi entropies. Phys. Lett. \textbf{198}: 119-121


\bibitem{Renyi}
R\'enyi A \textit{Probability Theory} (North-Holland, Amsterdam 1970)

\bibitem{Wada02}
Wada T (2002) On the thermodynamic stability conditions of Tsallis' entropy.
Phys. Lett. \textbf{297}: 334-337



\bibitem{Hotta99}
Hotta M and Joichi I (1999) Composability and generalized entropy
Phys. Lett. A \textbf{262} 302

\bibitem{Abe01}
Abe S (2001) General pseudoadditivity of composable entropy prescribed 
by the existence of equilibrium.
Phys. Rev. E \textbf{63}: 061105

\bibitem{Martinez01}
Mart\'{\i}nez S, Pennini F, Plastino A (2001)
Thermodynamics' zeroth law in a nonextensive scenario
Physica A \textbf{295}: 416-424

\bibitem{Abe01-Heat}
Abe S (2001)
Heat and entropy in nonextensive thermodynamics: transmutation from Tsallis theory to R\'enyi-entropy-based theory
Physica A \textbf{300}: 417-423

\bibitem{Wang03}
Wang QA, Le M\'ehaut\'e A (2003)
Unnormalized nonextensive expectation value and zeroth law of thermodynamics
Chaos, Solitons \& Fractals \textbf{15}: 537-541


\bibitem{Giorgio01}
Kaniadakis G (2001) Non-linear kinetics underlying generalized statistics.
Physica A \textbf{296}: 405-425

\bibitem{Giorgio-Antonio02}
Kaniadakis G, Scarfone AM (2002)
A new one-parameter deformation of the exponential function
Phisica A \textbf{305}: 69-75

\bibitem{Giorgio02}
Kaniadakis G (2002) Statistical mechanics in the context of special relativity.
Phys. Rev. E \textbf{66}: 056125

\end{thebibliography}
\end{document}